\documentstyle[prl,aps]{revtex}
\def\AmS{{\protect\the\textfont2
        A\kern-.1667em\lower.5ex\hbox{M}\kern-.125emS}}

\makeatletter
\tighten
\begin{document}
\draft
\title{The Anderson-Mott transition as a random-field problem}
\author{T.R.Kirkpatrick}
\address{Institute for Physical Science and Technology,\\
University of Maryland,\\
College Park, MD 20742}
\author{D.Belitz}
\address{Department of Physics and Materials Science Institute,\\
University of Oregon,\\
Eugene, OR 97403}

\date{\today}
\maketitle
\begin{abstract}
The Anderson-Mott transition of disordered interacting electrons is
shown to share many physical and technical features with classical
random-field systems. A renormalization group study of an order parameter
field theory for the Anderson-Mott transition shows that
random-field terms appear at one-loop order. They lead to an upper
critical dimension $d_{c}^{+}=6$ for this model. For $d>6$ the
critical behavior is mean-field like. For $d<6$ an $\epsilon$-expansion
yields exponents that coincide with those for the random-field Ising
model. Implications of these results are discussed.
\end{abstract}

\narrowtext

\par
It is a well established result that electrons in a random potential
at zero temperature undergo a metal-insulator transition as a function of
the disorder, provided that the space dimension $d>2$\cite{Anderson}. This
is true for both noninteracting \cite{LeeRama} and interacting \cite{R}
electrons, but the respective transitions are quite different in nature.
For the Anderson transition of noninteracting electrons no simple order
parameter (OP) description,
no upper critical dimension, and no Landau theory are known.
The Anderson-Mott transition (AMT) of interacting electrons,
on the other hand, has recently been shown to be conceptually simpler than
the Anderson transition in that it allows for a simple OP description with
the density of states (DOS) at the Fermi level as the OP.
In this picture the metal (insulator), with a nonzero (vanishing) DOS at the
Fermi level, represents the ordered (disordered) phase, respectively.
This OP description
leads to a finite upper critical dimension $d_{c}^{+}$, and to a Landau
theory of the AMT \cite{ourLetter}.

\par
This existence of an OP description raises important questions about the
nature of the fluctuations that drive the AMT, and about its relation to
other phase transitions in random systems. Consider a static, spin
independent random potential $u({\bf x})$ that couples to the electron
density. In a fermionic field theory this gives rise to a term in the
action \cite{R}
\begin{equation}
\int d{\bf x}\, u({\bf x}) \sum_{n} \bar\psi_{n}({\bf x}) \psi_{n}({\bf x})
\qquad,
\label{eq:1}
\end{equation}
with $\bar\psi$ and $\psi$ Grassmann fields, and $n$ a Matsubara
frequency index. Since the expectation value $\langle \bar\psi_{n}({\bf x})
\psi_{n}({\bf x}) \rangle$ determines the DOS, this means that the random
potential couples to the OP for the AMT. Magnetic transition where a
random field (RF) couples to the OP are known to have peculiar properties:
The RF fluctuations are dominant over the thermal fluctuations, which leads
to $d_{c}^{+}=6$ (rather than $4$) \cite{ImryMa}, and to a violation of
hyperscaling
even for $d<6$ \cite{Grinstein}. An $\epsilon$-expansion of the critical
exponents about $d=6$ leads to the famous 'dimensional reduction' problem
\cite{dimred}, and the critical dynamics have been proposed to be anomalous
\cite{FisherVillain}. It is then natural to ask whether similar phenomena
are to be expected at the AMT. Physically it is plausible that
interacting disordered electron systems should display RF effects since
they have the same type of frustration that occurs in RF magnets:
The random potential favors a local electron arrangement that
conflicts with the one favored by the electron-electron interaction.

\par
In this Letter we show that there is indeed a close analogy between the
AMT and RF problems. Within a renormalization group (RG) treatment of
the AMT we find that RF type terms
appear which lead to $d_{c}^{+}=6$. A $6-\epsilon$ expansion to first order
in $\epsilon$ leads to critical exponents that are identical with those of
the RF Ising model, and hyperscaling is violated due to a dangerous irrelevant
variable.

\par
Our starting point is the nonlinear $\sigma$-model description of interacting
disordered electrons \cite{F}. This is a Gaussian field
theory for a hermitian matrix field $\tilde{Q}({\bf x})$ with constraints
$(\tilde{Q}({\bf x}))^2 = \openone$,
with $\openone$ the unit matrix, and $tr\, \tilde{Q}({\bf x})=0$. $\tilde{Q}$
is a classical field comprising two fermionic fields. It carries two
Matsubara frequency indices $n,m$ and two replica
indices $\alpha,\beta$ (quenched disorder
has been incorporated by means of the replica trick).  The matrix elements
$\tilde{Q}^{\alpha\beta}_{nm}(\bf{x})$ are in general spin
quaternions, with the quaternion degrees of freedom describing the
particle-hole and particle-particle channel, respectively. For the sake of
simplicity, in this paper we restrict ourselves to the particle-hole
spin-singlet degrees of freedom, although the general model can be treated in
the same way \cite{ustbp}. The matrix elements of $\tilde{Q}$ can then
be expanded as
$\tilde{Q}_{nm}^{\alpha\beta} = \sum_{r=0,3}\>{_{r}\tilde{Q}_{nm}^
{\alpha\beta}}
\,\tau_{r}$ with $\tau_{0,1,2,3}$ the quaternion basis. We write the
action in the form
\widetext
\begin{equation}
S[\tilde{Q},\Lambda]=\frac{-1}{2G}\int d{\bf x}\ tr
\biggl[\Lambda({\bf x})[\tilde{Q}^2({\bf x})-
\openone]+\Bigl(\partial_{\bf x}\tilde{Q}({\bf x})\Bigr)^2\biggr]+2H \int
d{\bf x}\ tr\Bigl(\Omega \tilde{Q}({\bf x})\Bigr)
-\frac{\pi T}{4} K_s[\tilde{Q}({\bf x})\circ \tilde{Q}({\bf
x})]\qquad  .
\label{eq:2}
\end{equation}
\narrowtext
\noindent
Here $G$ is a measure of the disorder, $\Omega$ is a diagonal matrix whose
elements are the Matsubara frequencies $\omega_n$, and $H$ is
proportional to the free electron DOS.
$K_{s}<0$ is a repulsive electron-electron
interaction coupling constant in the particle-hole spin-singlet channel.
We consider a short-ranged model interaction for simplicity.
$[\tilde{Q}\circ\tilde{Q}]$ denotes a product in
frequency space which is given explicitly in
Refs.\ \cite{F,R}. Notice that we have enforced the constraint
$\tilde{Q}^{2}=\openone$ by means of an auxiliary, or ghost, matrix field
$\Lambda({\bf x})$.

\par
The correlation functions of $\tilde{Q}$ determine the
physical quantities.
Correlations of the $\tilde{Q}_{nm}$ with
$nm<0$ determine the diffusive modes which describe charge and spin
diffusion, while the DOS is determined by
$\langle\tilde{Q}^{\alpha\alpha}_{nn}\rangle$\cite{F,R}.
It is therefore convenient to separate $\tilde{Q}$ into blocks:
\begin{eqnarray}
\tilde{Q}^{\alpha\beta}_{nm}=
\Theta(nm)Q^{\alpha\beta}_{nm}({\bf x})+\Theta(n)\Theta(-
m)q^{\alpha\beta}_{nm}({\bf x})
\nonumber\\
+\Theta(-n)\Theta(m)(q^+)^{\alpha\beta}_{nm}({\bf x}).
\label{eq:3}
\end{eqnarray}
We then integrate out the massless $q$-field. Since the action,
Eq.(\ref{eq:2}), is quadratic in $q$ this can be done exactly.

\par
We next expand $Q$ and $\Lambda$ about their expectation values
$\langle{_{r} Q_{12}({\bf x})}\rangle \equiv \delta_{r0} \delta_{12} N_{n_{1}}$
and $\langle{_{r} \Lambda_{12}({\bf x})}\rangle \equiv
\delta_{r0} \delta_{12} l_{n_{1}}$, with $1 \equiv (n_{1},\alpha_{1})$, etc.,
\begin{mathletters}
\label{eqs:4}
\begin{equation}
{_{r} Q_{12}}({\bf x}) = \delta_{r0}\,\delta_{12}\, N_{n_{1}} +
{_{r}\phi_{12}}({\bf x})\qquad,
\label{eq:4a}
\end{equation}
\begin{equation}
{_{r} \Lambda_{12}}({\bf x}) = \delta_{r0}\,\delta_{12}\, l_{n_{1}} +
{_{r}\psi_{12}}({\bf x})\qquad,
\label{eq:4b}
\end{equation}
\end{mathletters}%
with $\langle {_{r}\phi_{12}}({\bf x})\rangle = \langle {_{r}\psi_{12}}
({\bf x})\rangle = 0$. Notice that $N_{n}$ is proportional to the DOS
at an energy $\omega_{n}$ measured from the Fermi surface \cite{R}.
The resulting action is quadratic in $\phi$, but contains terms with
arbitrary powers in $\psi$. If one formally integrates out $\psi$ one
is left with an action in terms of the OP $Q$ only. It can be shown that
the terms of higher than second order in $\psi$ do not change the
structure of the resulting $\phi$-field theory \cite{ustbp}. It is
therefore sufficient to integrate out $\psi$ in Gaussian approximation.
This is easily accomplished with the result
\widetext
\begin{mathletters}
\label{eqs:5}
\begin{eqnarray}
S[\phi] = -\int d{\bf x}\> tr\biggl[\phi ({\bf x})
\Bigl(-\partial_{{\bf x}}^{2}+\langle \Lambda\rangle\Bigr)\phi ({\bf x})
+{u\over G}\Bigl(\bigl(\langle Q\rangle \phi({\bf x})\bigr)^{2}
+\langle Q\rangle^{2}\phi^{2}({\bf x})\Bigr)\biggr]
+ {\Delta\over 2}\int d{\bf x}\, \sum_{i=>,<}
\Bigl(htr_{i}\, \phi({\bf x})\Bigr)^{2}
\nonumber\\
- u\int d{\bf x}\> tr\,\phi^{4}({\bf x})
- {2u\over \sqrt{2G}}\int d{\bf x}\>
tr\,\Bigl[\langle Q\rangle \phi^{3}({\bf x}) +
\phi({\bf x})\langle Q \rangle \phi^{2}({\bf x})\Bigr]
\nonumber\\
- {u\over G}\int d{\bf x}\> tr\,
\biggl[A \Bigl(\phi^{2}({\bf x}) + {2\over \sqrt{2G}}\langle Q\rangle
\phi({\bf x}) \Bigr)\biggr]
- {2\over \sqrt{2G}}\int\! d{\bf x}\> tr\, \Bigl[B\phi ({\bf x})\Bigr] \quad,
\label{eq:5a}
\end{eqnarray}
where we have scaled $\phi$ by a factor of $\sqrt{2G}$.
$tr \equiv \sum_{r,\alpha,n}$ denotes a trace over all discrete
degrees of freedom, and
$htr_{>,<} \equiv \sum_{r,\alpha}\sum_{n\ge 0,n<0}$
denotes 'half-traces' that
extend only over positive and negative frequencies, respectively.
$A$ and $B$ are functions of
$\langle Q\rangle$ and $\langle\Lambda\rangle$, and are given by
\begin{equation}
A(\langle Q\rangle,\langle\Lambda\rangle) =
\langle Q\rangle^{2} - 1 + f\bigl(\langle \Lambda
\rangle\bigr)\qquad,\qquad
B(\langle Q\rangle,\langle\Lambda\rangle) =
\langle \Lambda\rangle\langle Q \rangle - 2GH\Omega\qquad,
\label{eq:5b}
\end{equation}
where $f(\langle \Lambda\rangle)$
is a matrix with elements ${_{r}f_{12}} = \delta_{r0}\,\delta_{12}\,
f_{n_{1}}$ with
\begin{equation}
f_{n}=- \frac{G}{4}\sum_{\bf p}\sum^{-{\infty}}_{m=-1}\frac{2\pi
TGK_s}{[p^2+\frac{1}{2}(l_{n} +l_{m})]^2}
\left[1+\sum^{n-m-1}_{n_{1}=0}\frac{2 \pi TGK_s}{p^2 +
\frac{1}{2}(l_{n_{1}} +l_{n_{1} - n+m})}
\right]^{-1}\qquad,
\label{eq:5c}
\end{equation}
\end{mathletters}%
\narrowtext
\noindent
for $n\ge 0$, and a similar expression for $n<0$.
The bare value of the coupling constant $u$ is $u=-G/(df/dl)$. The bare
value of $\Delta$ is zero. However, a term of this structure is generated
by the RG at one-loop order, and it is crucial to include it in the
action. A comparison with Refs.\ \cite{ImryMa,Grinstein} shows that the
half-trace
terms have the characteristic replica structure of a RF term. Since it is
quadratic in $\phi$ this term contributes to the Gaussian (G) propagator.
In the replica limit we find
\begin{eqnarray}
\Bigl<{_{r}\phi_{12}}({\bf k})\,{_{s}\phi_{34}}({\bf p})\Bigr>^{(G)} =
\delta({\bf k} + {\bf p})\,\delta_{rs}\,{1\over 16}{1\over k^{2}+m_{12}}
\nonumber\\
\times \biggl[\delta_{13}\,\delta_{24} + (-)^{r}\,\delta_{14}\,\delta_{23}
+ {4\Delta\,\Theta(n_{1}n_{3})\over k^{2}+m_{12}}
\,\delta_{r0}\,\delta_{12}\,\delta_{34}\biggr]\,\,.
\label{eq:6}
\end{eqnarray}
Here $m_{12} \equiv (l_{1}+l_{2})/2+u(N_{1}+N_{2})^{2}$.
It is clear that the term proportional to $\Delta$
will increase the upper critical dimension by $2$, as it does in the
case of RF magnets.
We therefore expect $d_{c}^{+}=6$ in the model
given by Eqs.\ (\ref{eqs:5}) rather than $d_{c}^{+}=4$ which one would
conclude from a power-counting analysis of the action at zero-loop
order, i.e. without the RF term \cite{ourLetter,dcplusfootnote}.

\par
Equation (\ref{eq:5a}) requires some explanatory comments.
(1) $\langle Q\rangle$ and $\langle\Lambda\rangle$ are
determined by the conditions $\langle\phi\rangle=0$ and $\langle\psi
\rangle=0$. At zero-loop
order, these two conditions yield $A\bigl(\langle Q\rangle,
\langle\Lambda\rangle\bigr) = B\bigl(\langle Q\rangle,\langle\Lambda
\rangle\bigr)=0$. This is the zero-loop order equation of state
that has been discussed
in Ref.\ \cite{ourLetter}. It yields
mean-field exponents, which constitute the exact critical behavior
for $d>d_{c}^{+}=6$. For $d<6$, renormalizations of the equation of
state change the critical behavior. (2) In writing Eq.\ (\ref{eq:5a})
we have omitted some terms that are irrelevant by power counting for
$d>4$. So are the terms of $O(\phi^{4})$ which we kept.
However, as we will see, the latter couple to the
RF coupling constant $\Delta$ and therefore {\it must} be kept. We have
verified that none of the terms omitted,
and no other terms generated by the RG, couple to $\Delta$
\cite {ustbp}. These considerations are in direct analogy to the case of
a magnet in a RF \cite{ImryMa}.

\par
We now perform a one-loop RG analysis of the action, Eq.\ (\ref{eq:5a}),
using a standard momentum-shell method \cite{WilsonKogut}. It is convenient
to first consider the theory at criticality. Then we can put $\langle Q\rangle
= \langle\Lambda\rangle=0$, and consider the renormalization of $u$ for
$d=6-\epsilon$. Since $u$ is irrelevant for $d>4$
we need to keep only contributions that
are of order $g\equiv u\Delta$, $\Delta$ being relevant with a
bare dimension of $2$. $\Delta$ is not renormalized,
and the $\partial_{{\bf x}}^{2}$-term is not renormalized either to
one-loop order, so
the exponent $\eta=O(\epsilon^2)$. We obtain
the following flow equation for $g$,
\begin{equation}
{dg \over d\ln b} = \epsilon g - {9\over 2}g^2 + O(g^3)\quad,
\label{eq:7}
\end{equation}
with $b$ the RG length rescaling factor. Eq.\ (\ref{eq:7})
possesses a fixed point
$g^{*} = 2\epsilon/9 + O(\epsilon^{2})$.

\par
We now turn to the disordered phase, i.e. the insulator where
$N_{n=0}$ vanishes and $l_{n=0}\equiv l$ has a
nonzero value. We thus put $\langle Q\rangle =0$, and renormalize the mass
term $\langle\Lambda\rangle$ or $l$ in the action.
We obtain
\begin{equation}
{d l\over d\ln b} = 2l - gl + O(g^{2})\quad,
\label{eq:8}
\end{equation}
and two equations that determine the renormalized equation of state,
\begin{mathletters}
\label{eqs:9}
\begin{equation}
N^{2} = 1 - f(l) - {Gg \over 4u}\sum_{{\bf p}}{1 \over (p^{2} + l)^{2}}
\quad,
\label{eq:9a}
\end{equation}
\begin{equation}
lN = 2GH\Omega - {gN\over 2}\sum_{{\bf p}}{1 \over (p^{2}+l)^{2}}\quad,
\label{eq:9b}
\end{equation}
\end{mathletters}%
where both $N$ and $l$ are to be considered as functions of $\Omega$.
If we replace $g$ in Eq.\ (\ref{eq:8}) by its fixed point value $g^{*}$,
we find that $l$ scales like
\begin{equation}
l(b) \sim b^{2[1-\epsilon/9+O(\epsilon^{2})]}\quad.
\label{eq:10}
\end{equation}
We next invoke the equation
of state to find the relation between $l$ and the distance from the critical
point, $t$. To this end, we expand the
r.h.s. of Eq.\ (\ref{eq:9a}) for small values of $l$. The $l$-independent
contribution is $t$. At linear order
in $d=6$ one finds a term $\sim l$, and a term
$\sim l\ln l$. The prefactors of these two terms are related since
$df/dl = -G/u$. Replacing $g$ by
$g^{*}$, we can exponentiate and find,
\begin{equation}
l \sim t^{1+\epsilon/18+O(\epsilon^{2})}\quad.
\label{eq:11}
\end{equation}
This holds for $\epsilon >0$. For $d>6$ one finds instead $l \sim t$ as
one would expect within mean-field theory.

\par
We can now combine Eqs.\ (\ref{eq:10}) and (\ref{eq:11}) to get the correlation
length exponent $\nu$ to first order in $\epsilon$. Since we also know
$\eta$ to this order, standard scaling arguments yield all other static
exponents. We find
\widetext
\begin{equation}
\nu = {1\over 2} + {\epsilon\over 12} + O(\epsilon^{2})\,,\,
\eta = 0 + O(\epsilon^{2})\,,\,
\gamma = 1 + {\epsilon\over 6} + O(\epsilon^{2})\,,\,
\beta = {1\over 2} - {\epsilon\over 6} + O(\epsilon^{2})\,,\,
\delta = 3 + \epsilon + O(\epsilon^{2})\,.
\label{eqs:12}
\end{equation}
\narrowtext
\noindent
In order to obtain $\beta$ and $\delta$ we have used the fact that $u$ is
dangerously irrelevant, so hyperscaling is violated (see \cite{Grinstein},
and the discussion below), and have accordingly
replaced $d$ by $d-2$ in all $d$-dependent scaling laws. Notice that
Eqs.\ (\ref{eqs:12}) are identical with the corresponding results for
a RF Ising model \cite{ImryMa,Aharony}.

\par
We still need to determine the dynamical scaling exponent $z$. For this
purpose we obtain a relation between $l$, $N$, and $\Omega$ from Eq.\
(\ref{eq:9b}). Expanding the r.h.s. for small $l$, going to criticality,
and exponentiating, we find
$Nl^{1+\epsilon/9+O(\epsilon^{2})} \sim \Omega$.
If we combine this with Eq.\ (\ref{eq:11}), in which we substitute
$t \sim N^{1/\beta}$, then we find
\begin{equation}
z = 3 - \epsilon/2 + O(\epsilon^{2})\quad.
\label{eq:14}
\end{equation}
Notice that $z=\delta\beta/\nu=y_{h}$, with $y_{h}$ the exponent of the field
conjugate to the OP. This was to be expected since our RG did not
involve any frequency integrals.

\par
Let us discuss our results.
Grinstein \cite{Grinstein} has shown that hyperscaling is violated
in RF magnets because because the quartic coupling constant
is dangerously irrelevant. The same
arguments apply here. In order to completely describe static scaling, we
therefore need a third exponent, $\theta$, in addition to the usual two
independent exponents $\nu$ and $\eta$. $\theta$ describes the flow of $u$
to zero, and enters all hyperscaling relations. Consider, for instance,
the OP. Its scale dimension is $d/2-1+\eta/2$, which in the absence of
dangerous irrelevant variables leads to the scaling law $\beta=\nu(d-2+\eta)
/2$. However, since $u$ scales to zero like $u \sim b^{-\theta}$, and since
$N(u\rightarrow 0) \sim 1/\sqrt{u}$,
one has instead
\begin{equation}
\beta = \nu (d-2-\theta+\eta)/2\quad.
\label{eq:15}
\end{equation}
Our explicit one-loop calculation yields $\theta = 2 + O(\epsilon^{2})$,
but we expect $\theta = 2$ to all orders in $\epsilon$ as is the case in
the RF Ising model \cite{ParisiSourlas}. Now consider the density
susceptibility, $\partial n/\partial \mu$, or the specific heat coefficient,
$\gamma = \lim_{T\rightarrow 0}C_{V}/T$, or the spin susceptibility,
$\chi_{s}$. All scale like an inverse volume
times a time, so their naive scale dimension is $d-z$. The violation of
hyperscaling changes this to $d-\theta -z$. Above we have seen that $z$
is equal to $y_{h}=\delta \beta/\nu$, and
therefore
\begin{mathletters}
\label{eqs:16}
\begin{equation}
z=(d-\theta+2-\eta)/2\quad.
\label{eq:16a}
\end{equation}
Somewhat surprisingly for a quantum phase transition, the dynamics are not
independent from the statics. This is due to the RF fluctuations being
stronger than the quantum fluctuations.
We thus find that all thermodynamic susceptibilities scale like the OP, viz.
\begin{equation}
\chi(t,\Omega) = b^{(2+\theta-d-\eta)/2}\,\chi(tb^{1/\nu},\Omega b^{z})
\quad,
\label{eq:16b}
\end{equation}
\end{mathletters}%
where $\chi$ can stand for $N$, $\partial n/\partial\mu$, $\gamma$, or
$\chi_{s}$.
$\Omega$ can stand for either external frequency, or temperature, or, in
the case of $N$, for the distance from the Fermi level.

\par
RF problems contain an anomalously divergent correlation function
\cite{Grinstein}. In the present case this function describes 'mesoscopic'
fluctuations of the local DOS, $C({\bf x},{\bf y}) =
[N({\bf x})N({\bf y})]_{av} - N^{2}$, with $N({\bf x})$ the (unaveraged)
local DOS at the Fermi level, and $[...]_{av}$ the ensemble average. At
criticality $C$ behaves like \cite{Grinstein} $C(k\rightarrow 0)\sim
k^{-2+\eta-\theta}$. An experimental observation of such a strong divergence
would indicate that RF features are indeed present at the AMT.

\par
We now consider transport properties. The charge or spin diffusivity $D$ has
a scale dimension of $z-2$. Since no $d$ is involved, $\theta$ does not
enter, and we have
\begin{equation}
D(t,\Omega) = b^{2-z}\,D(tb^{1/\nu},\Omega b^{z})\quad.
\label{eq:17}
\end{equation}
One is also interested in the scaling behavior of the conductivity,
$\sigma = D\partial n/\partial\mu$. The behavior of
$\sigma$ depends on whether or not $\partial n/\partial\mu$
has an analytic background contribution in addition to the critical
contribution given by Eq.\ (\ref{eq:16b}). Ref.\ \cite{ourLetter}
has argued that for the present model it does not. This yields
\begin{mathletters}
\begin{equation}
\sigma(t,\Omega) = b^{2+\theta-d}\,\sigma(tb^{1/\nu},\Omega b^{z})\quad.
\label{eq:18a}
\end{equation}
The conductivity exponent $s$ is then
\begin{equation}
s=\nu(d-2-\theta)\quad.
\label{eq:18b}
\end{equation}
In more general models
$\partial n/\partial\mu$ might have a noncritical background contribution.
In that case $\sigma$ will scale like the diffusivity, leading to
\begin{equation}
s=\nu(z-2)={\nu\over 2}(d-2-\theta-\eta)\quad.
\label{eq:18c}
\end{equation}
\end{mathletters}%
In either case, Wegner scaling (i.e., $s=\nu(d-2)$ \cite{Wegner76}), which
in previous work in $d=2+\epsilon$
had been found to hold for the AMT as well as for the Anderson
transition \cite{R}, is violated. This removes
the requirement $s\ge 2(d-2)/d$, which follows from Wegner scaling combined
with the result of Ref.\ \cite{Chayes}, and has led to severe problems with
the interpretation of certain experiments \cite{R}.

\par
Finally, we note that our results imply that all of the complications (most
of them not quite understood) that are known to occur in the RF magnet
problem should be expected for the AMT as well. For
instance, although $\theta=2$ to all orders in perturbation theory, this
almost certainly changes due to nonperturbative effects \cite{dimred}.
$d=4$ is most likely some sort of a critical dimension.
Finally, the non-power law dynamical critical
behavior that has been proposed for RF magnets \cite{FisherVillain}
should be expected to occur at the AMT as well, giving the AMT some aspects
of a glass transition, despite the conventional
power law scaling encountered in perturbation theory.

\par
It is a pleasure to thank John Toner for discussions.
This work was supported by the NSF under grant numbers DMR-92-17496
and DMR-92-09879.


\begin{references}
\bibitem{Anderson} P.~W. Anderson, Phys. Rev. {\bf 109}, 1492 (1958);
 E. Abrahams, P.~W. Anderson, D.~C. Licciardello, and T.~V. Ramakrishnan,
 Phys. Rev. Lett. {\bf 42}, 673 (1979).
\bibitem{LeeRama} For a review, see, P.~A. Lee and T.~V. Ramakrishnan, Rev.
 Mod. Phys. {\bf 57}, 287 (1985).
\bibitem{R} For a review, see, e.g., D. Belitz and T.~R. Kirkpatrick,
 Rev. Mod. Phys. {\bf 66}, 261 (1994).
\bibitem{ourLetter} T.~R. Kirkpatrick and D. Belitz, Phys. Rev. Lett. {\bf 73},
 xxxx (1994).
\bibitem{ImryMa} Y. Imry and S.~K. Ma, Phys. Rev. Lett. {\bf 35}, 1399 (1975);
 for a review see, e.g., T. Nattermann and J. Villain, Phase Transitions
 {\bf 11}, 5 (1988).
\bibitem{Grinstein} G. Grinstein, Phys. Rev. Lett. {\bf 37}, 944 (1976).
\bibitem{dimred} See, e.g., D.~S. Fisher, Phys. Rev. B {\bf 31}, 7233 (1985),
 and references therein.
\bibitem{FisherVillain} D.~S. Fisher, Phys. Rev. Lett. {\bf 56}, 416 (1986);
 J. Villain, J. Phys. (Paris) {\bf 46}, 1843 (1985).
\bibitem{F} A.~M. Finkel'stein, Zh. Eksp. Teor. Fiz. {\bf 84}, 168 (1983)
 [JETP {\bf 57}, 97 (1983)]. This is a generalization of Wegner's model,
 F. Wegner, Z. Phys. B {\bf 35}, 207 (1979).
\bibitem{ustbp} D. Belitz and T.~R. Kirkpatrick, unpublished.
\bibitem{dcplusfootnote} See Ref.\ 10 in Ref.\ \cite{ourLetter}, which
 anticipated the possibility that $d_{c}^{+}$ might be larger than
 power counting at zero-loop order suggests. See also A.~B. Harris and
 T.~C. Lubensky, Phys. Rev. B {\bf 23}, 2640 (1981). Our treatment of
 the AMT owes much to this failed attempt of an OP description of the
 Anderson transition.
\bibitem{WilsonKogut} K.~J. Wilson and J. Kogut, Phys. Rep. {\bf 12}, 75
 (1974).
\bibitem{Aharony} Since the RF couples only to the 'longitudinal' field,
 $Q_{nn}$, the problem is structurally similar to the anisotropic
 RF magnetic model considered by A. Aharony, Phys. Rev. B {\bf 18}, 3328
 (1978), which also yielded Ising exponents. As a check of our RG procedures
 we have rewritten that model as a nonlinear $\sigma$-model, performed the
 same RG calculation, and obtained Aharony's results.
\bibitem{ParisiSourlas} A. Aharony, Y. Imry, and S.~K. Ma, Phys. Rev. Lett.
 {\bf 37}, 1364 (1976), G. Parisi and N. Sourlas, Phys. Rev. Lett. {\bf 43},
 744 (1979).
\bibitem{Wegner76} F. Wegner, Z. Phys. B {\bf 25}, 327 (1976).
\bibitem{Chayes} J. Chayes, L. Chayes, D.~S. Fisher, and T. Spencer, Phys.
 Rev. Lett. {\bf 57}, 2999 (1986).
\end{references}
\end{document}